\documentclass[12pt]{article} 
\usepackage[koi8-r]{inputenc}
\usepackage[english]{babel}
\usepackage{pazh}
\usepackage{graphicx,amssymb,amsmath}
\usepackage{natbib}
\usepackage{xspace}
\tightenlines

\newcommand{\Teff}{$T_{\rm eff}$}
\newcommand{\lgg}{log\,$g$}
\newcommand{\feh}{[Fe/H]\xspace}
\newcommand{\eps}[1]{\log\varepsilon_{\rm #1}}
\newcommand{\kms}{km\,s$^{-1}$}
\newcommand{\kH}{$S_{\!\rm H}$}    
\newcommand{\Eexc}{$E_{\rm exc}$}

\newcommand{\eu}[5]{\mbox{$#1\,^#2{\rm #3}^{#4}_{\rm #5}$}}

\sloppypar

\begin{document}
	
	\baselineskip 21pt

	\title{\bf Scandium abundances of F-G-K dwarf stars in a wide metallicity range}

\author{\bf \hspace{-1.3cm}\copyright\, 2022 \ \
L.~Mashonkina\affilmark{1*}, A.~Romanovskaya\affilmark{1} }

\affil{
		{\it Institute of Astronomy, Russian Academy of Sciences, Pyatnitskaya st. 48, 119017 Moscow, Russia$^1$} }

	\vspace{2mm}
	
	\sloppypar 
	\vspace{2mm}

A new model atom of \ion{Sc}{2} was constructed using the most up-to-date atomic data. For the testing purpose, the non-local thermodynamic equilibrium (non-LTE) calculations were carried out for three stars with reliably determined atmospheric parameters: the Sun, HD~61421 (Procyon), and HD~84937. Accounting for deviations from LTE leads to smaller abundance errors compared with the LTE case and consistent within the error bars abundances obtained from different \ion{Sc}{2} lines. Solar non-LTE abundance $\eps{Sun} = 3.12\pm0.05$ exceeds the meteoritic abundance recommended by Lodders (2021), by 0.08~dex. But agreement within 0.02~dex with the meteoritic abundance is obtained for Procyon. Using high-resolution spectra, we determined the scandium LTE and non-LTE abundances for 56 stars in the  metallicity range $-2.62 \le$ [Fe/H] $\le$ 0.24. The dependence of [Sc/Fe] on [Fe/H] demonstrates a similarity with the behavior of the $\alpha$-process elements: scandium is enhanced relative to iron ([Sc/Fe] $\sim$ 0.2) for [Fe/H] $< -1$, and [Sc/Fe] decreases with increasing [Fe/H] for the higher metallicity. There is a hint of a tight relation between abundances of scandium and titanium. The results obtained provide observational constraints to the scenarios of scandium origin.


\vfill
\noindent\rule{8cm}{1pt}\\
{$^*$ e-mail $<$lima@inasan.ru$>$}

\section{Introduction}

There are several problems associated with the stellar scandium abundances.

The determinations of the Sc abundance in the solar atmosphere give values that differ from each other by more than 2$\sigma$: from $\eps{Sun}$ = 3.07$\pm$0.04 \citep{2008A&A...481..489Z} to $\eps{Sun}$ = 3.16$\pm$0.04 \citep{scott_sc} and exceed the meteoritic abundance $\eps{met}$ = 3.04$\pm$0.03 \citep{lodders21}. Hereafter, an abundance scale is used in which $\eps{}$(H) = 12.

Scandium exhibits different behavior in A-type stars with iron abundances close to solar ([Fe/H]\footnote{for any two elements X and Y: [X/Y] = log$(N_{\rm X}/N_{\rm Y})_{star} - \log (N_{\rm X}/N_{\rm Y})_{Sun}$.} $\sim$ 0, normal A stars), and non-magnetic A stars with strong metal lines (Am stars, [Fe/H] $\succsim$ 0.5). But Sc does not belong to the group of metals with strong lines. On the contrary, Am stars reveal a deficit of Sc, which can reach an order of magnitude (Adelman et al. 2000 and references in their article). Therefore, the scandium abundance plays a key role in classifying a star as Am. However, the reasons for the different behavior of Sc and elements of the Fe group in A-type stars remain unclear.

An origin of Sc remains unclear. According to the most recent models of the chemical evolution of the Galaxy \citep{Kobayashi2020}, the main source of Sc during the life of the Galaxy were type II supernovae (SNeII) and hypernovae (HN), but they produced scandium an order of magnitude less than is observed in the Galactic matter. The observational data on the Sc abundance of stars in a wide range of metallicity serves to test the assumptions made in the Galactic chemical evolution model and establish constraints to the model parameters. In the literature, we find determinations of the Sc abundance in large samples of stars with very low Fe abundances ([Fe/H] $< -2$), for example, works by \cite{Cayrel2004}, \cite{Yong_sc}, \cite{Roederer_sc}, but there are very few data for stars with [Fe/H] $> -2$. Zhao et al. (2016) obtained the Sc abundance for 49 dwarf stars in the range $-2.6 \le$ [Fe/H] $\le 0.25$ and Reggiani et al. (2017) for 23 stars with [Fe/H] between $-2.8$ and $-1.5$.

In this paper, two of the mentioned problems are considered, namely, those concerning the Sun and the Galactic trend [Sc/Fe] -- [Fe/H]. The problem of scandium in normal A stars and Am stars will be the subject of our next article. We determine the solar Sc abundance from lines of \ion{Sc}{2} using the most up-to-date atomic data and a line-formation modeling free from the simplifying assumption of local thermodynamic equilibrium (LTE). It is referred to as the non-LTE approach. It is important to find an answer to the question of whether the solar atmosphere Sc abundance represents the cosmic one or whether there are physical processes that change the Sc abundance of the Sun in comparison with the meteoritic abundance.

Second, we revise the Sc abundance of the sample of stars from Zhao et al. (2016) using a new model of the scandium atom and abandoning the differential approach. Zhao et al. (2016) determined [Sc/H] = $\eps{star} - \eps{Sun}$ for each individual line in order to exclude the influence of errors in the oscillator strengths (or $gf$-values) on the final result. This was a forced approach due to the lack of precise atomic parameters of the lines, but it excluded from the analysis the \ion{Sc}{2} lines in
range 4246-4420~\AA\ due to strong blending effects in the solar spectrum. Exactly these lines are most reliably measured in stars with [Fe/H] $< -2$. As a result, for very metal-poor (VMP, [Fe/H] $< -2$) stars, Zhao et al. (2016) were able to use only two to three lines, and the Sc abundance was not determine for two VMP stars. In our work, we used laboratory measurements of $gf$ from Lawler et al. (2019). Therefore, for each star, the abundance was determined for all observed lines of \ion{Sc}{2}. To improve statistics in the range $-1.5 <$ [Fe/H] $< -1$, the stellar sample was complemented by five stars from \cite{mash2003}.

To solve the tasks posed, a model atom of \ion{Sc}{2} was constructed using the transition probabilities from Lawler et al. (2019) and the electron collision rates from Grive and Ramsbottom (2012). Earlier, the non-LTE method for \ion{Sc}{1}--\ion{Sc}{2} was developed by Zhang et al. (2008) and applied by Zhao et al. (2016) to determine stellar non-LTE abundances.

The paper is structured as follows. The new model atom is presented in Section~\ref{sect:atom}. It is tested in Section~\ref{sect:sun} by analyzing the \ion{Sc}{2} lines in the spectra of stars with reliably determined atmospheric parameters, that is the Sun, HD~61421 (Procyon), and HD~84937. In Section~\ref{sect:trend}, we determine the scandium LTE and non-LTE abundances of the sample stars and analyze the resulting trend [Sc/Fe] -- [Fe/H]. Finally, we summarize our conclusions and recommendations.

\section{New model of the scandium atom}\label{sect:atom}

\subsection{Atomic data}

{\it Energy levels.} The atom model includes three lower terms of \ion{Sc}{1}, 888 energy levels of \ion{Sc}{2} and the ground state of \ion{Sc}{3}. The level energies are taken from the NIST  database\footnote{https://www.nist.gov/pml/atomic-spectra-database} (Kramida et al. 2019) and the atomic structure calculations by Kurucz (2009)\footnote{\tt http://kurucz.harvard.edu/atoms/2101/}. With ionization energy $\chi$ = 6.54~eV, scandium is strongly ionized in the atmospheres of stars with an effective temperature of \Teff\ $>$ 4500~K, and a number density of \ion{Sc}{1} depends strongly on the accuracy of calculations of the ionization and recombination rates. In the absence of accurate photoionization cross sections and collisional processes, we exclude the possibility of an accurate calculation of the statistical equilibrium (SE) of \ion{Sc}{1}. For the constraint of particle conservation, it suffices to have the three most populated lower terms of \ion{Sc}{1}.

\begin{figure*}  
	\centering
	\includegraphics[width=0.80\columnwidth,clip]{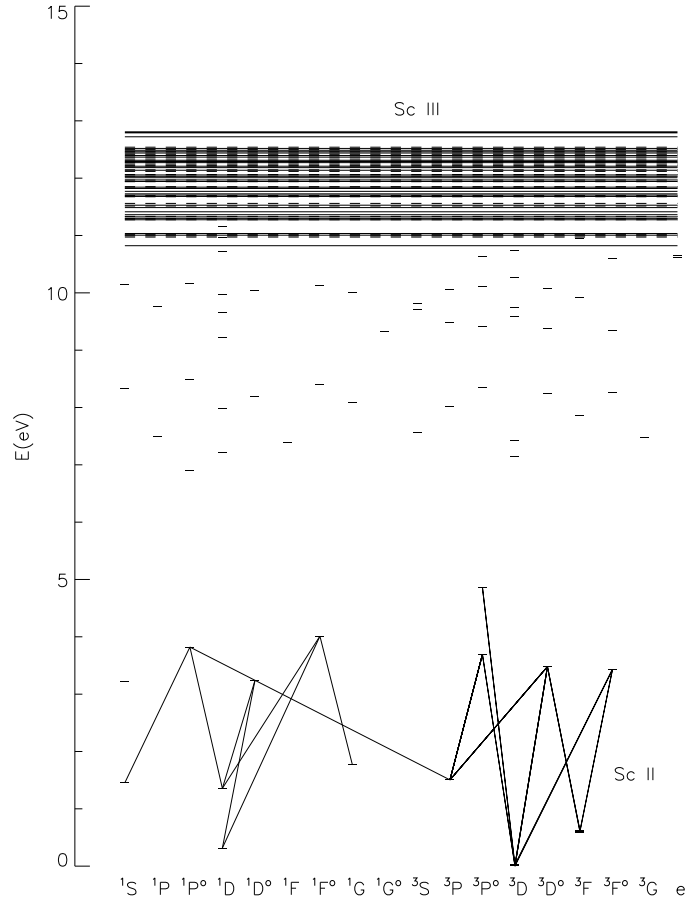}
	\caption{\ion{Sc}{2} levels, included in the model atom, and the transitions corresponding to the spectral lines, used in abundance determinations. Solid and dashed horizontal lines indicate the energy of even and odd superlevels, respectively. Levels with violation of the LS-coupling are presented in column 'e'.}
	\label{fig:atom_Sc2}
\end{figure*}

The fine splitting is taken into account for the \ion{Sc}{2} \eu{3d4s}{3}{D}{}{} ground state and the low-excitation term \eu{3d^2}{3}{F}{}{}. High-excitation levels of \ion{Sc}{2} with a small energy separation and of the same parity were combined. Such superlevels were constructed from the levels predicted in the atomic structure calculations, but not discovered (so far) in lab measurements. The average energy of the combined superlevel was calculated taking into account the statistical weights of individual levels. The final model atom includes 79 even and 55 odd levels of \ion{Sc}{2}. The highest levels of \ion{Sc}{2} are separated from the ground state of \ion{Sc}{3} by 0.08--0.26~eV, which is much lower than the average kinetic energy of electrons at temperatures of up to 20\,000 K. This ensures effective coupling of the \ion{Sc}{2} levels with the ground state of \ion{Sc}{3} through collisions. The term diagram is shown in Fig.~\ref{fig:atom_Sc2}.

{\it Radiative rates.} The model atom includes 4032 allowed bound-bound (b-b) transitions. $gf$-values are taken from Lawler et al. (2019) and Kurucz (2009). For the \ion{Sc}{2} transitions, which are associated with the low-excitation levels and in which the upper levels can be pumped by ultraviolet (UV) radiation, the radiative rates are calculated using the Voigt function for the absorption profile. For other transitions, the Doppler absorption profile is adopted.
The photoionization cross sections are calculated in the hydrogenic approximation using the effective principal quantum number instead of the principal quantum number. Note that the number density of \ion{Sc}{2} practically does not depend on the accuracy of calculations of the ionization/recombination rates, since \ion{Sc}{2} is the dominant ionization stage in the atmospheres of the studied objects.

{\it Collision rates.} In the atmospheres of late spectral type stars, the number density of electrons is much lower than the number density of neutral hydrogen atoms; therefore, the excitation of levels and the formation of ions can occur as a result of collisions not only with electrons, but also with \ion{H}{1} atoms. For electron-impact excitation, we use data from \cite{sc2_e} obtained with the R-matrix method. They are available for 948 b-b transitions of \ion{Sc}{2} between levels with an excitation energy of \Eexc $\le$ 9.5~eV. Note that all the observed lines of \ion{Sc}{2} are formed between levels in this energy interval. For the remaining b-b transitions, electronic collisions are calculated using the van Regemorter (1962) formula if the transition is allowed, and an effective collision strength is assumed to be 1 if the transition is forbidden. Electron-impact ionization rates are calculated using Seaton's (1962) formula and the adopted photoionization cross sections at the  threshold.

To account for collisions with \ion{H}{1}, leading to excitation of \ion{Sc}{1} and \ion{Sc}{2}, we use the \cite{1984A&A...130..319S} formula. Since the formula is approximate, with an accuracy of only the order of magnitude, the calculations were made with several values of the scaling factor \kH\ = 0, 0.1 and 1.

\begin{figure*}  
	\centering
	\includegraphics[width=0.45\columnwidth,clip]{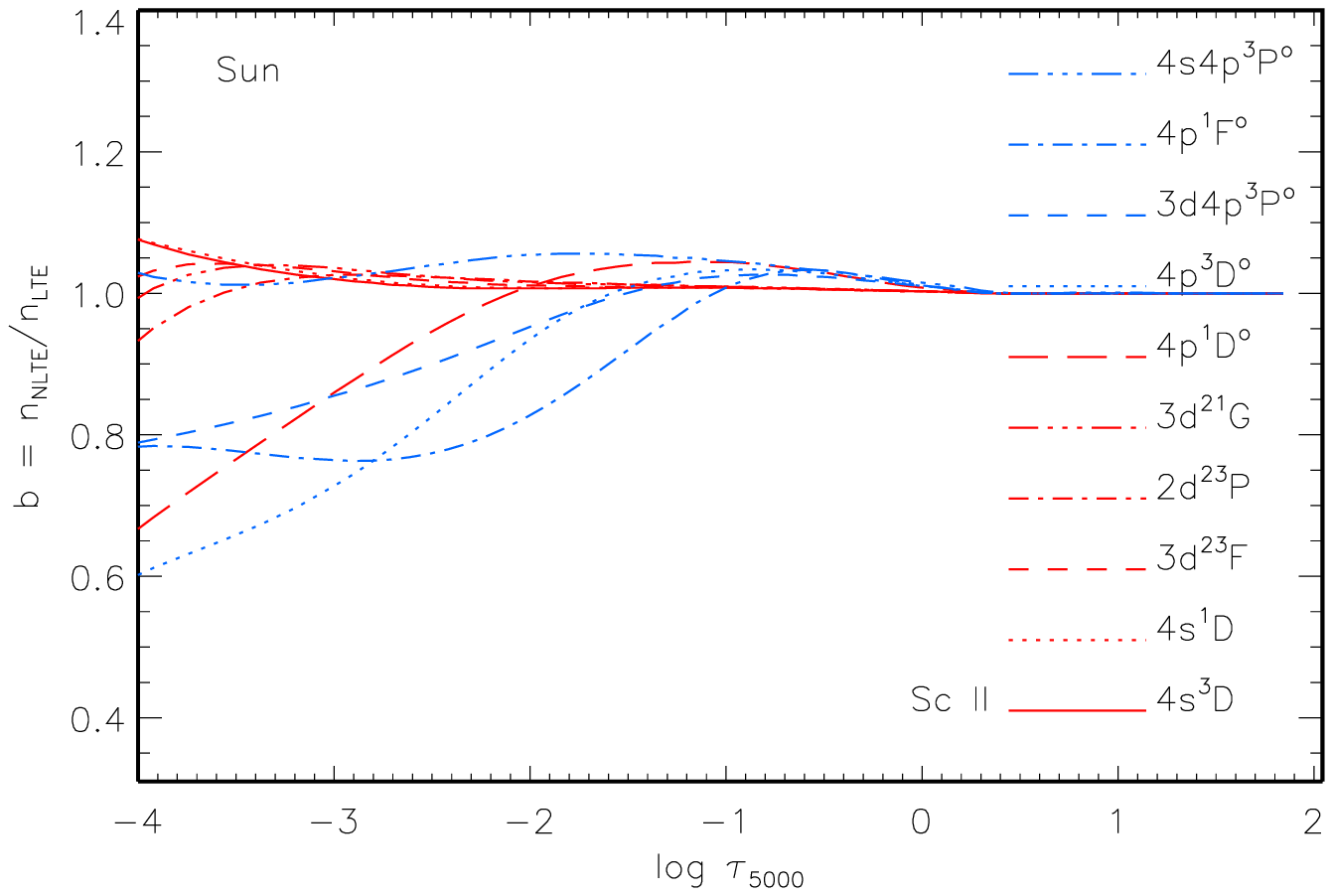}
	\includegraphics[width=0.45\columnwidth,clip]{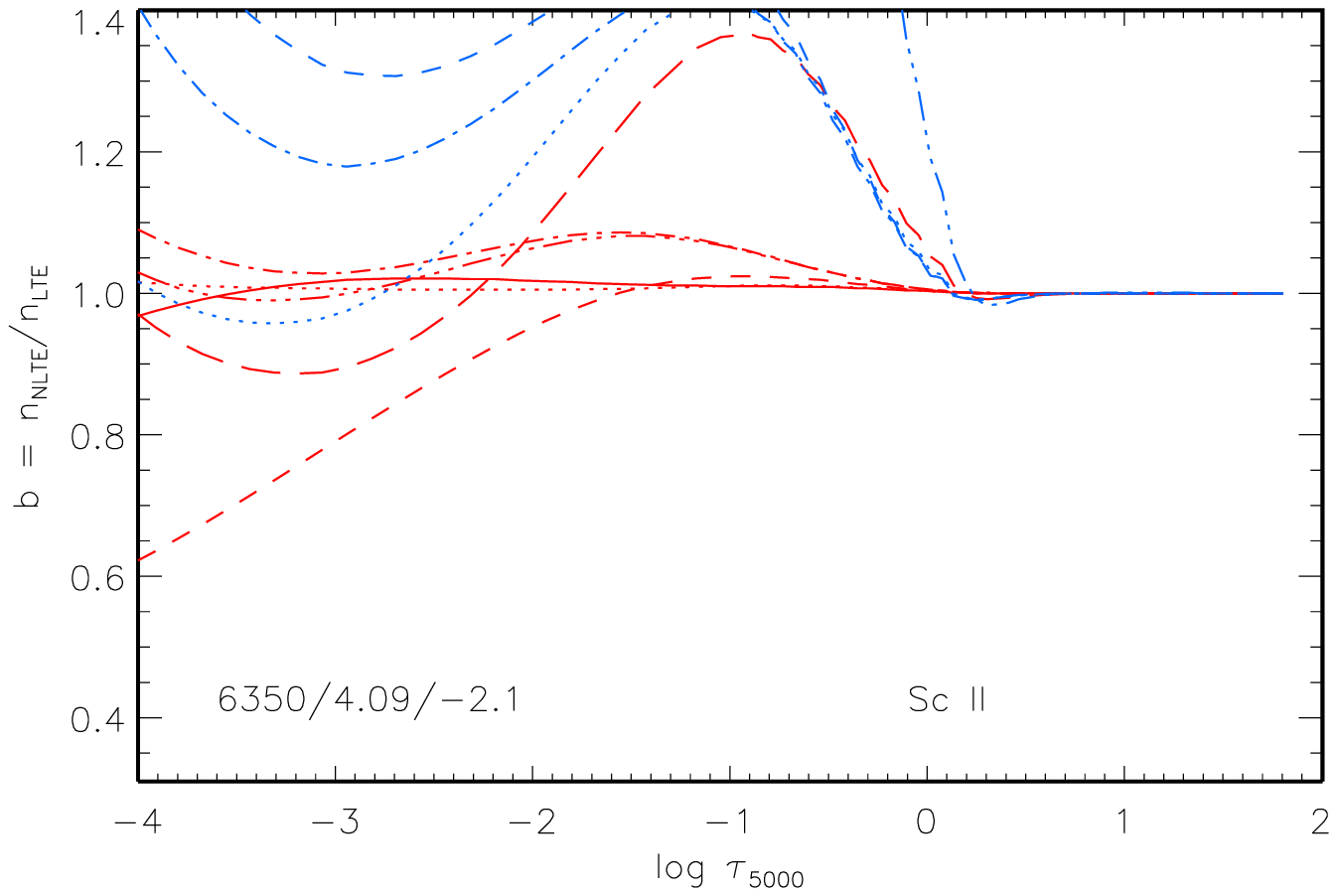}
	\caption{b-factors of the selected levels of \ion{Sc}{2} in the model atmospheres 5780/4.44/0 and 6350/4.09/$-2.1$.}
	\label{fig:bf}
\end{figure*}

\subsection{Statistical equilibrium of \ion{Sc}{2} in stellar atmospheres of different metallicity}\label{sect:nlte}

The system of the statistical equilibrium equations and radiative transfer in a given atmospheric model is solved using the code DETAIL developed by \cite{giddings81} and \cite{butler84} based on the accelerated $\Lambda$-iteration method \citep{RH91,RH92}. The opacity package was updated by T. Gehren, J. Reetz, L. Mashonkina, as described in \cite{mash_fe}.

Throughout this work, we use plane-parallel (1D) atmospheric models from the MARCS database\footnote{http://marcs.astro.uu.se} (Gustafsson et al. 2008). The models with given \Teff, surface gravity \lgg\ and \feh were obtained by interpolation using the algorithm posted on the MARCS website.

Figure~\ref{fig:bf} shows b-factors, b = $n_{\rm NLTE} /n_{\rm LTE}$, of individual levels of \ion{Sc}{2}, related to the observed lines, in the model atmospheres with solar metallicity (\Teff/\lgg/\feh = 5780/4.44/0) and with a large deficiency of metals (6350/4.09/$-2.1$). Here, $n_{\rm NLTE}$ and $n_{\rm LTE}$ are the level populations obtained by solving the SE equations (non-LTE) and by the Boltzmann-Saha formulas (LTE). Since in both models \ion{Sc}{2} is the dominant ionization stage, then the populations of its ground state \eu{3d4s}{3}{D}{}{} and the first excited level \eu{3d4s}{1}{D}{}{} (\Eexc\ = 0.3~eV) are close to the LTE populations (b $\simeq$ 1) everywhere in the atmosphere. Since the intensity of UV radiation is higher in a hotter and metal-poor atmosphere, the radiative pumping of the odd terms \eu{3d4p}{3}{F}{\circ}{}, \eu{3d4p}{3}{P}{\circ}{} and \eu{4s4p}{3}{P}{\circ}{} via transitions from the ground state (lines 3613-3642~\AA, 3361-3372~\AA\ and 2552-2563~\AA) is more efficient in the 6350/4.09/$-2.1$ model. Therefore the behavior of all levels above \Eexc\ = 0.3~eV differs in the two models. In the 6350/4.09/$-2.1$ model, already in deep layers, the medium becomes optical thin for radiation in the spontaneous transitions from overpopulated (b $>$ 1) odd terms to the low-excitation levels, which makes the populations of the latter also higher than the LTE  ones, although to a lesser extent than those of the upper levels. An example is the transition \eu{3d^2}{3}{P}{}{} -- \eu{3d4p}{3}{P}{\circ}{}, in which the 5641~\AA\ line is formed.

\section{\ion{Sc}{2} lines in spectra of the Sun, Procyon, and HD~84937}\label{sect:sun}

In order to test the new model atom, we analyzed lines of \ion{Sc}{2} in the three reference stars for which there are high-resolution spectra and the atmospheric parameters are reliably determined. These are the Sun, the star HD~61421 (Procyon, F5 IV-V) with a metallicity close to the solar one, and a VMP  star HD~84937.

\subsection{Observational material and atmospheric parameters}\label{sect:obs}

We analyze the spectrum of the Sun as a star using the atlas of Kurucz et al. (1984). Spectral resolving power is R = $\lambda/\Delta\lambda \simeq$ 300000. As in our previous studies, we use the well-known parameters of the solar atmosphere: \Teff\ = 5780~K, \lgg\ = 4.44, and a microturbulence velocity of $\xi_{t}$ = 0.9~\kms.

Spectrum of Procyon obtained with R $\simeq$ 80\,000 was taken from the UVESPOP archive \citep{Bagnulo2003}. Atmosphere parameters \Teff\ = 6615~K, \lgg\ = 3.89, \feh = $-0.05$, and $\xi_{t}$ = 2~\kms\ were determined by \cite{Ryabchikova2016} by automatically fitting the theoretical spectrum to the selected regions in the same observed spectrum. Note that \Teff\ = 6615~K agrees within the determination error (89~K) with the effective temperature obtained by \cite{Boyajian2013} in the range from 6562~K to 6597~K based on interferometric measurements of Procyon's angular diameter.

The observed spectrum (R $\simeq$ 80\,000) HD~84937 is also taken from the UVESPOP archive \citep{Bagnulo2003}. In addition, we use the ultraviolet spectrum obtained on the Hubble Space Telescope with the STIS spectrograph in the range of 1875-3158\,\AA, with a resolution of R $\simeq$ 25\,000. This data is provided by Thomas Ayres on the ASTRAL project website\footnote{\tt http://casa.colorado.edu/$\sim$ayres/ASTRAL/}. Atmospheric parameters \Teff\ = 6350~K, \lgg\ = 4.09, [Fe/H] = $-2.16$, and $\xi_{t}$ = 1.7~\kms\ were determined by \cite{lickI} using spectroscopic and photometric methods and a known distance.

\subsection{Determination of scandium abundances}

\begin{table}
	\centering
	\renewcommand{\tabcolsep}{3pt}
	\caption{Atomic parameters of the \ion{Sc}{2} lines and the LTE and non-LTE (NLTE, \kH\ = 0.1) abundances of scandium ($\eps{}$) for the reference stars. }
	\vspace{3mm}
	\label{tab:list}
	\begin{tabular}{lcrcccccc}\hline
	$\lambda$, \AA  & \Eexc, & log $gf$ & \multicolumn{2}{c}{Sun} & \multicolumn{2}{c}{Procyon} & \multicolumn{2}{c}{HD~84937} \\
	                &   eV   &          & \multicolumn{2}{c}{5780/4.44/0} & \multicolumn{2}{c}{6615/3.89/$-0.01$} & \multicolumn{2}{c}{6350/4.09/$-2.16$} \\
	                &        &          &  ~LTE  & NLTE~~ & ~LTE  & NLTE~~ &  ~LTE  & NLTE  \\
		\noalign{\smallskip} \hline \noalign{\smallskip}
2552.35$^1$ & 0.022 &  0.05 &  -   &  -   &  -   &  -   & 0.99 & 1.23 \\
2563.19 & 0.000 & --0.57 &  -   &  -   &  -   &  -   & 0.98 & 1.15 \\
3353.72 & 0.315 &  0.26 &  -   &  -   &  -   &  -   & 1.03 & 1.13 \\
3359.68 & 0.008 & --0.75 &  -   &  -   &  -   &  -   & 1.09 & 1.16 \\
3368.94 & 0.008 & --0.39 &  -   &  -   &  -   &  -   & 1.09 & 1.18 \\
3567.70 & 0.000 & --0.47 &  -   &  -   &  -   &  -   & 1.07 & 1.15 \\
3576.34 & 0.008 &  0.01 &  -   &  -   &  -   &  -   & 1.01 & 1.10 \\
3580.92 & 0.000 & --0.14 &  -   &  -   &  -   &  -   & 1.07 & 1.16 \\
3589.63 & 0.008 & --0.57 &  -   &  -   &  -   &  -   & 1.10 & 1.17 \\
3590.47 & 0.022 & --0.55 &  -   &  -   &  -   &  -   & 1.11 & 1.17 \\
3613.83 & 0.022 &  0.42 &  -   &  -   &  -   &  -   & 1.07 & 1.11 \\
3642.78 & 0.000 &  0.05 &  -   &  -   &  -   &  -   & 1.05 & 1.13 \\
3645.31 & 0.022 & --0.41 &  -   &  -   &  -   &  -   & 1.13 & 1.18 \\
4246.82 & 0.315 &  0.24 &  -   &  -   &  -   &  -   & 1.14 & 1.22 \\
4314.08 & 0.618 & --0.11 & 3.22 & 3.15 & 3.17 & 3.05 & 1.15 & 1.27 \\
4320.73 & 0.605 & --0.28 &  -   &  -   &  -   &  -   & 1.16 & 1.27 \\
4325.00 & 0.595 & --0.44 &  -   &  -   &  -   &  -   & 1.14 & 1.25 \\
4400.38 & 0.605 & --0.54 & 3.18 & 3.16 & 3.13 & 3.07 & 1.17 & 1.26 \\
4420.66 & 0.618 & --2.33 & 3.18 & 3.18 & 3.08 & 3.07 &  -   &  -   \\
4670.40 & 1.357 & --0.60 & 3.14 & 3.11 & 3.12 & 3.08 & 1.11 & 1.20 \\
5031.02 & 1.357 & --0.41 &  -   &  -   &  -   &  -   & 1.16 & 1.26 \\
5239.81 & 1.455 & --0.76 & 3.13 & 3.13 & 3.08 & 3.09 &  -   &  -   \\
5357.20$^{1,2}$ & 1.500 & --2.11 & 3.15 & 3.15 & 3.04 & 3.06 &  -   &  -   \\
5526.77 & 1.768 & --0.01 & 3.24 & 3.16 & 3.21 & 3.09 & 1.17 & 1.29 \\
5640.99 & 1.500 & --0.99 & 3.14 & 3.13 & 3.08 & 3.06 &  -   &  -   \\
5657.88 & 1.507 & --0.54 & 3.20 & 3.18 & 3.14 & 3.09 & 1.08 & 1.24 \\
5658.35 & 1.497 & --1.17 & 3.15 & 3.15 &  -   &  -   &  -   &  -   \\
5669.04 & 1.500 & --1.10 & 3.11 & 3.11 &  -   &  -   &  -   &  -   \\
5684.19 & 1.507 & --1.03 & 3.15 & 3.14 & 3.09 & 3.07 &  -   &  -   \\
6245.62$^3$ & 1.500 & --1.02 & 3.04 & 3.03 & 3.01 & 2.99 &  -   &  -   \\
6279.74 & 1.500 & --1.33 & 3.14 & 3.13 &  -   &  -   &  -   &  -   \\
6300.68$^3$ & 1.510 & --1.90 & 2.99 & 3.00 & 3.04 & 3.02 &  -   &  -   \\
6320.83$^3$ & 1.500 & --1.82 & 3.06 & 3.05 & 3.04 & 3.03 &  -   &  -   \\
6604.58 & 1.357 & --1.26 & 3.08 & 3.09 & 2.99 & 3.00 &  -   &  -   \\
\noalign{\smallskip}\hline \noalign{\smallskip}
\multicolumn{3}{l}{Mean} & 3.14 & 3.12 & 3.09 & 3.06 & 1.09 & 1.19 \\
\multicolumn{3}{l}{$\sigma$(dex)} & 0.06 & 0.05 & 0.06 & 0.03 & 0.06 & 0.06 \\
\noalign{\smallskip}\hline \noalign{\smallskip}
\multicolumn{9}{l}{$^1$ no HFS; $^2$ $gf$ \citep{LD89}; $^3$ $gf$ \citep{K09}.}
	    \end{tabular}
\end{table}

{\it Lines of \ion{Sc}{2}}, used in abundance determinations, together with their atomic parameters are listed in Table~\ref{tab:list}. We use oscillator strengths based on laboratory measurements of \cite{lawler2019}. Exceptions are the \ion{Sc}{2} 5357~\AA\ line with log$gf$ measured by Lawler and Dakin (1989) and the lines at 6245, 6300 and 6320~\AA, for which there are only calculations by Kurucz (2009).

The levels in scandium atoms and ions are subject to hyperfine splitting (HFS). Therefore, in calculating the theoretical line profiles, it is necessary to take into account their HFS structure. For all studied lines, except for \ion{Sc}{2} 2552~\AA\ and 5357~\AA, wavelengths and oscillator strengths of the HFS components are available in the updated VALD database (Pakhomov et al. 2019), as well as on the page {\tt https://github.com/vmplacco/linemake}. We note the complete agreement of the data in these two sources. In the solar spectrum, the \ion{Sc}{2} 5357~\AA\ line has an equivalent width of $EW \simeq$ 5~m\AA, and neglecting the HFS structure does not affect the theoretical $EW$ and derived abundance. \ion{Sc}{2} 2552~\AA\ was measured only for HD~84937. Possibly, due to neglecting the HFS structure, it gives an abundance that is 0.08~dex higher than the average for other UV lines.

Van der Waals damping constants, log~$\Gamma_6$, for lines of \ion{Sc}{2} were calculated by \cite{K09} and presented in the VALD database \citep{2015PhyS...90e4005R,vald_hfs}. They have close values for different lines and, per 10\,000~K, range from log~$\Gamma_6 = -7.81$ to log~$\Gamma_6 = -7.83$.

\begin{figure}  
	\centering
	\includegraphics[width=0.45\columnwidth,clip]{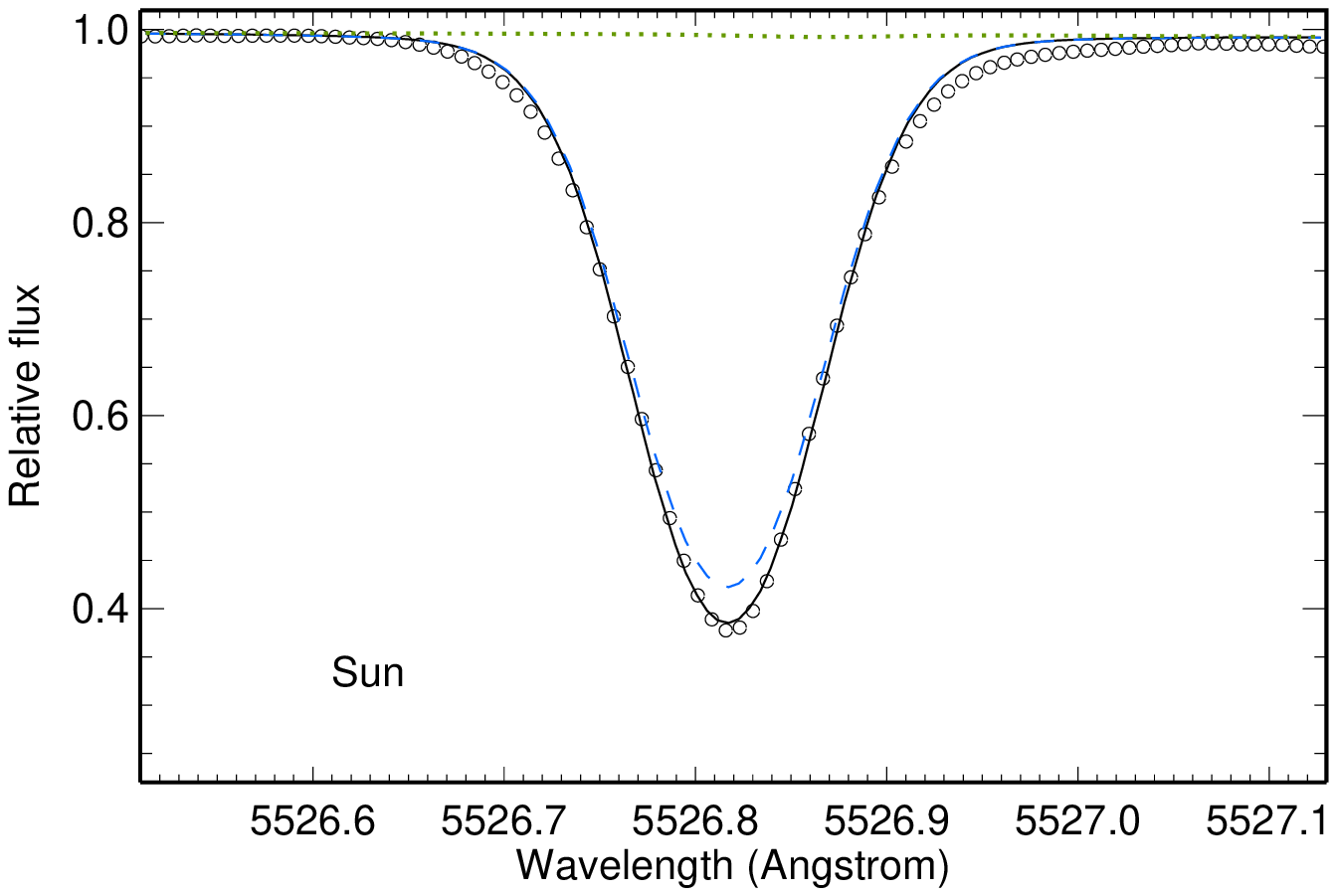}
	\includegraphics[width=0.45\columnwidth,clip]{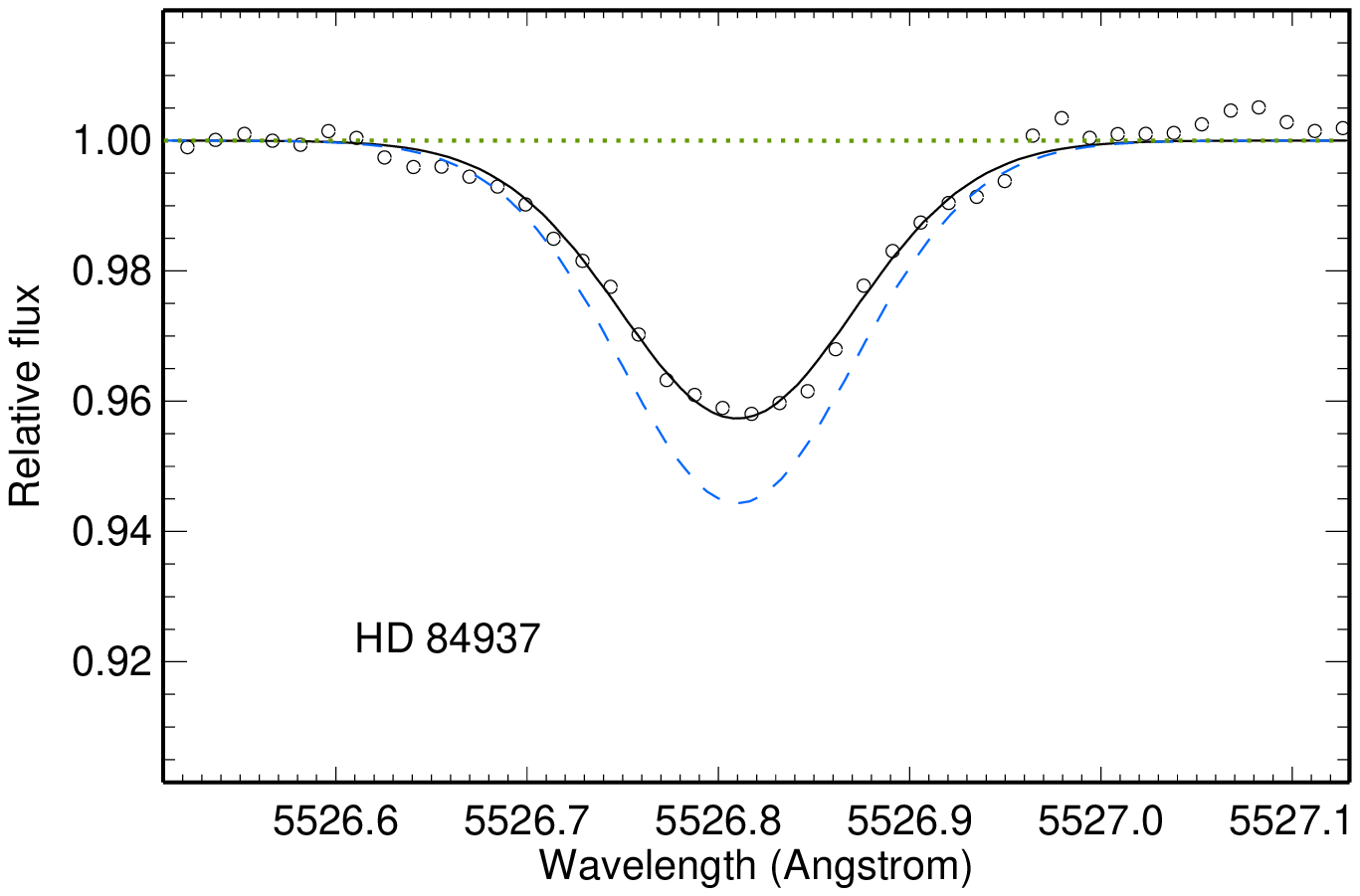}
	\caption{\ion{Sc}{2} 5526~\AA\ in the spectra of the Sun and HD~84937 (circles) and the best non-LTE fits (\kH\ = 0.1, solid curve). The dashed curves corresponds to the LTE calculations. For the Sun and HD~84937, calculations were made using $\eps{}$ = 3.16 and 1.29, respectively.}
	\label{fig:sc5526}
\end{figure}

{\it The codes and methodology for abundance determinations.} Everywhere in this study, the synthetic spectrum method is applied by automatically fitting the theoretical line profile to the observed one. Equivalent widths were measured in the same procedure only for the purpose of illustrating the results. We use the codes synthV\_NLTE (Tsymbal et al., 2019) and BinMag \citep{Kochukhov_binmag}. The list of lines for calculating the synthetic spectrum is taken from VALD. In addition to the model atmosphere, microturbulent velocity, and line atomic parameters, the spectral resolving power and the projection of the rotation velocity onto the line of sight $V\sin i$ were the fixed parameters in the procedure for matching the theoretical and observed spectra. For the Sun and Procyon, $V\sin i$ = 1.8~\kms\ and 4~\kms, respectively. As in \citet{Zhao2016}, rotational broadening was also taken into account for six stars with $V\sin i \ge$ 6~\kms: HD~58855 ($V\sin i$ = 10~\kms), HD~89744 (9~\kms), HD~92855 (10~\kms), HD~99984 (6~\kms), HD~100563 (10~\kms), HD~106516 (7~\kms). For the rest of the studied stars, line broadening due to rotation was not taken into account, since it is much smaller than the broadening due to macroturbulent motions. Macroturbulent velocity $V_{\rm mac}$, scandium abundance, and wavelength shift of the line were free parameters in the fitting procedure. The Doppler shift, due to the radial velocity of the star and the orbital motion of the Earth around the Sun, was in advance taken into account. Figure~\ref{fig:sc5526} displays the best non-LTE fits of \ion{Sc}{2} 5526~\AA\ in the Sun and HD~84937, as examples.

{\it Non-LTE effects on lines of \ion{Sc}{2}.} For comparison, Fig.~\ref{fig:sc5526} shows the LTE profiles calculated with the abundance obtained in the non-LTE calculations. For the Sun, departures from LTE lead to slight strengthening \ion{Sc}{2} 5526~\AA, and, on the contrary, to weakening for HD~84937. \ion{Sc}{2} 5526~\AA\ is formed in the transition \eu{3d^2}{1}{G}{}{} -- \eu{3d4p}{1}{F}{\circ}{}. In the solar atmosphere, its core is formed at depths with log~$\tau_{5000} \sim -1.6$, where a population of the lower level is close to the LTE one, while the upper level is underpopulated (Fig.~\ref{fig:bf}). As a result, the line source function is less than the Planck function at these depths, and the line is strengthened. The difference between non-LTE and LTE abundances, $\Delta_{\rm NLTE}$ = $\eps{NLTE} - \eps{LTE}$, which is called the non-LTE abundance correction, is $-0.08$~dex in this case. In the atmosphere of HD~84937, the core of \ion{Sc}{2} 5526~\AA\ is formed at depths with log~$\tau_{5000} \sim -0.1$, where the population of the lower level is close to the LTE one, while the upper level is overpopulated (Fig.~\ref{fig:bf}), which leads to a weakening of the line. The non-LTE correction is positive and equal to $\Delta_{\rm NLTE}$ = 0.12~dex.

\begin{figure}  
	\centering
	\includegraphics[width=0.32\columnwidth,clip]{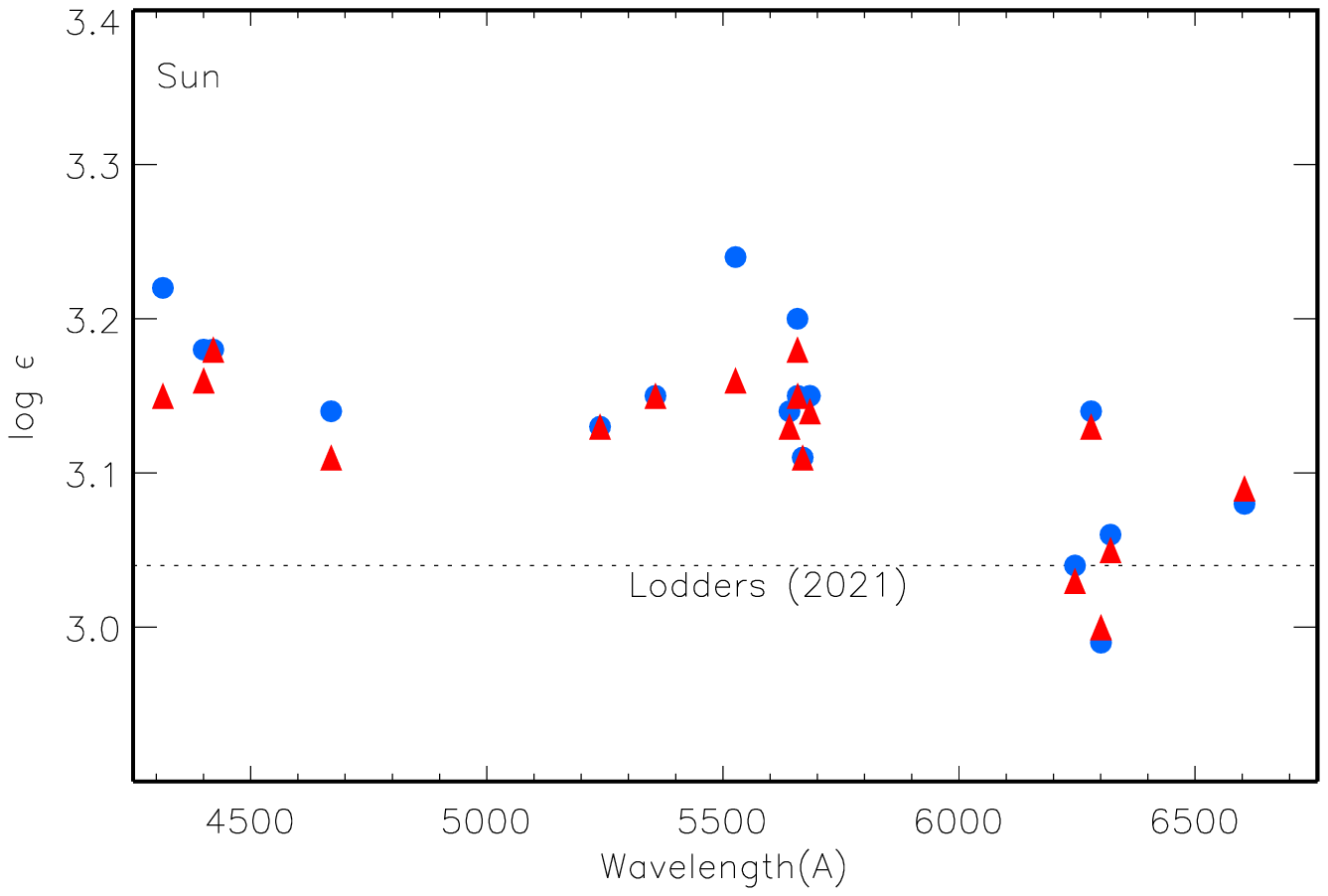}
	\includegraphics[width=0.32\columnwidth,clip]{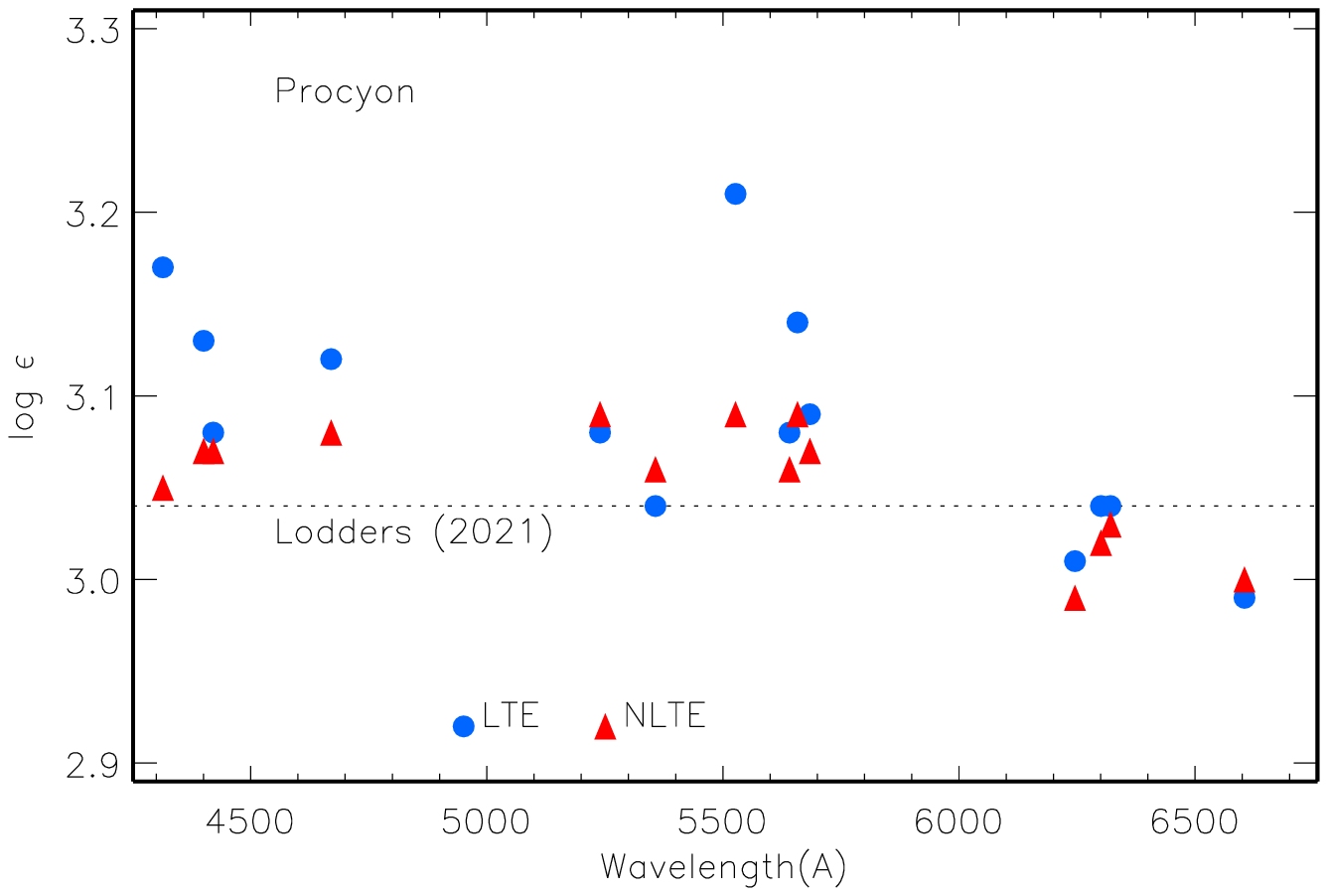}
	\includegraphics[width=0.32\columnwidth,clip]{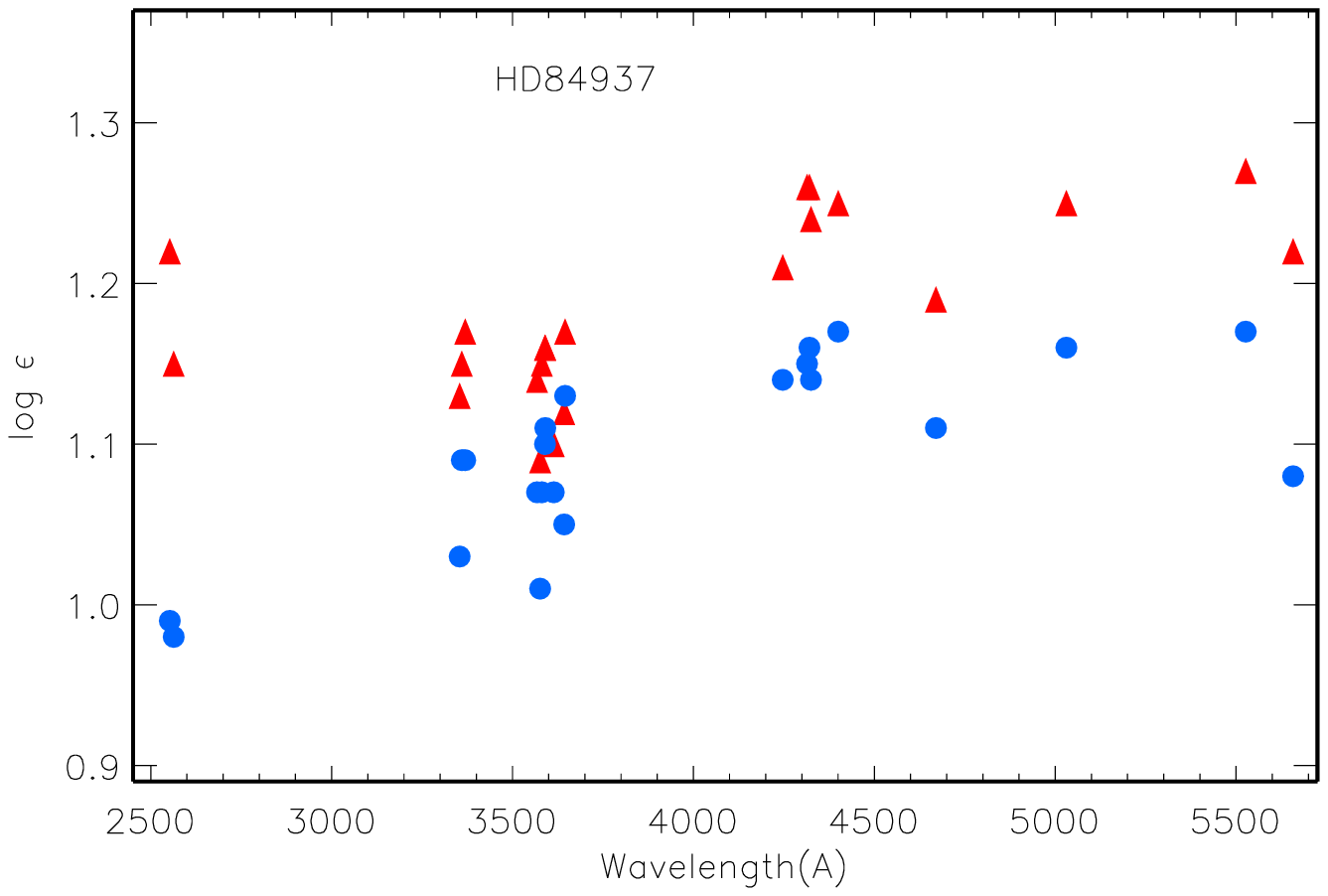}
	\caption{LTE (blue circles) and non-LTE (red triangles) abundances obtained from individual lines of \ion{Sc}{2} in Sun, Procyon and HD~84937. The dotted line denotes the meteoritic abundance from \cite{lodders21}.}
	\label{fig:abund_lines}
\end{figure}

\begin{figure}  
	\centering
	\includegraphics[width=0.45\columnwidth,clip]{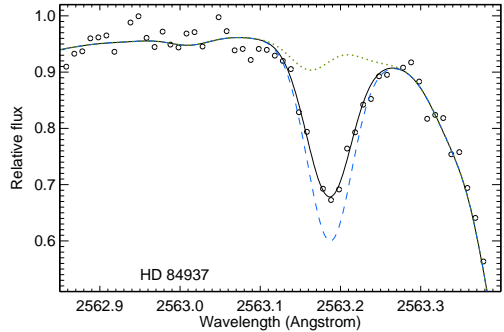}
	\caption{\ion{Sc}{2} 2563~\AA\ in the spectrum of HD~84937 (circles) and the best non-LTE fit (solid curve, $\eps{}$ = 1.15). The dashed curve corresponds to the LTE calculations with the same Sc abundance, and the dotted curve to the calculations in the absence of Sc in the atmosphere.}
	\label{fig:uv2563}
\end{figure}

The obtained LTE and non-LTE abundances from individual lines are presented in Table~\ref{tab:list} and in Fig.~\ref{fig:abund_lines} for each of the three reference stars. For the Sun, the non-LTE abundance corrections are negative and small in absolute value. Exception is \ion{Sc}{2} 6604~\AA\ with a small positive correction. Procyon is hotter, and the negative non-LTE corrections are larger in absolute value. HD~84937 reveals stronger non-LTE effects in the \ion{Sc}{2} lines, since the intensity of UV radiation is higher and the electron number density is lower, resulting in higher radiative rates and lower collisional rates. Lines of \ion{Sc}{2} are weaker in non-LTE than in LTE, and $\Delta_{\rm NLTE}$ are positive. For each of the stars non-LTE leads to a smaller dispersion in the single line measurements around the mean $\sigma = \sqrt{\sum(x-\bar{x})^2/(N_l - 1)}$, where $N_l$ is the number of lines. It is, in particular, well seen for Procyon, where $\sigma$ decreases by a factor of two, to 0.03~dex, and for HD~84937, where non-LTE removes the abundance difference  between the lines in the UV and visible regions. Figure~\ref{fig:uv2563} illustrates the reliability of abundance determination from \ion{Sc}{2} 2563~\AA\ in the spectrum of HD~84937. The \ion{Sc}{2}  line is located in the far wing of \ion{Fe}{1} 2563.399~\AA\ (\Eexc\ = 0.96~eV, log~$gf = -2.26$), which is well reproduced for given atomic parameters and iron abundance, and blends with \ion{Cr}{2} 2563.157~\AA\ (\Eexc\ = 6.805~eV, log~$gf = -1.683$), which is much weaker than \ion{Sc}{2} 2563~\AA\ and, due to the uncertainty in the Cr abundance (or $gf$), can affect the derived Sc abundance by no more than 0.01~dex.

{\it Influence of collisions with hydrogen atoms.} Non-LTE calculations were carried out for three variants of collisional rates in the SE equations: \kH\ = 0 -- pure electron collisions; \kH\ = 1 -- collisions with hydrogen atoms are accounted for with the formula of \cite{1984A&A...130..319S}; \kH\ = 0.1 -- the contribution of collisions with \ion{H}{1} is reduced by a factor of 10. For solar lines of \ion{Sc}{2}, the non-LTE corrections range from +0.01~dex to $-0.08$~dex, if \kH\ = 0, and decrease in absolute value by a maximum of 0.01~dex in the variant \kH\ = 1. Non-LTE effects are much stronger for HD~84937, with $\Delta_{\rm NLTE}$ from 0.04~dex to 0.24~dex in option \kH\ = 0. However, collisions with \ion{H}{1} only weakly affect the final results, namely, the difference in non-LTE abundances between variants \kH\ = 0 and 1 does not exceed 0.02~dex. Weak influence of collisions with hydrogen atoms on the magnitude of departures from LTE for the \ion{Sc}{2} lines is due to the \ion{Sc}{2} term structure and, possibly, the lack of exact data on the rates of these processes. The formula of \cite{1984A&A...130..319S} was designed only for allowed transitions, while all the \ion{Sc}{2} levels with \Eexc\ $<$ 2~eV have the same parity. For allowed transitions with energies above 3~eV, radiative processes predominate over collisional ones. Hereafter, a compromise value of \kH\ = 0.1 was adopted to obtain the final non-LTE abundances.

The resulting solar non-LTE abundance is 0.02~dex less than the LTE value and closer to the meteoritic abundance $\eps{met} = 3.04 \pm 0.03$ \citep{lodders21}, but still exceeds it by more than 1$\sigma$.  It should be noted that in different works the magnitude of meteoritic abundance varies within the error of determination. For example, \cite{Lodders2009} recommended $\eps{met} = 3.07 \pm 0.02$. In Procyon, the scandium non-LTE abundance is consistent with the meteoritic value.

\subsection{Comparison with the literature data}

\begin{figure}  
	\centering
	\includegraphics[width=0.45\columnwidth,clip]{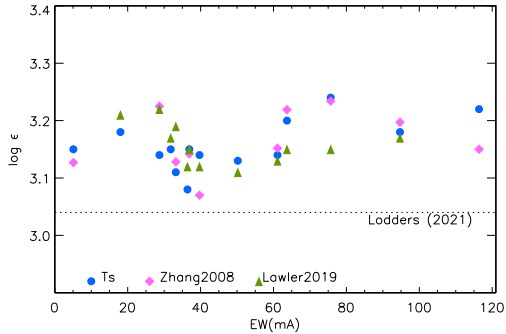}
	\caption{LTE abundances obtained from individual lines of \ion{Sc}{2} in the solar spectrum by different authors. Circles, diamonds, and triangles correspond to this study (Ts), \cite{2008A&A...481..489Z}, and \cite{lawler2019}, respectively. The data of \cite{2008A&A...481..489Z} were converted into the system of oscillator strengths by \cite{lawler2019}. The equivalent widths are from our measurements. The dotted line denotes the meteoritic abundance from \cite{lodders21}).}
	\label{fig:compare_sun}
\end{figure}

The obtained mean solar abundance agrees within the error bars with the data of other authors if the abundances from individual lines are reduced to one system of $gf$-values. Figure~\ref{fig:compare_sun} shows the LTE abundances from individual lines as derived by \cite{lawler2019}, \cite{2008A&A...481..489Z} and in this paper. Abundances of \cite{2008A&A...481..489Z} were recalculated using $gf$-values, measured by Lawler et al. (2019).
Differences are minimal between us and \cite{2008A&A...481..489Z}, since in both works the same Atlas of the Sun \citep{KPNO1984} was used and the theoretical model atmospheres, although from different sources, that is, from \cite{2008A&A...486..951G} and \cite{mafags04}, respectively. The mean abundances agree within 0.01~dex, however, the difference reaches nearly 0.1~dex for three lines (4314, 5640, 5658~\AA), which is most likely due to different treatment of blending lines.\cite{lawler2019} used the solar disc center spectrum \citep{sun_intensity1973} and a semi-empirical  model atmosphere \citep{HM74} and obtained the higher, on average, abundance than our LTE value, by 0.02~dex.

Now we discuss the original data. Using the oscillator strengths from the calculations of Kurucz, Zhang et al. (2008) obtained $\eps{}$(Z08)= 3.10$\pm 0.05$ and $3.07 \pm 0.04$ in LTE and non-LTE. This was the first and so far the only work on non-LTE analysis of the solar \ion{Sc}{2} lines. Despite the improved model atom, our calculations show similar non-LTE effects for \ion{Sc}{2}, because of their small magnitude.

\cite{scott_sc} determined the solar Sc abundance as $\eps{}$(3D) = 3.17$\pm 0.04$ based on the hydrodynamic LTE calculations with the 3D model atmosphere and using the non-LTE corrections published by Zhang et al. (2008). They used laboratory $gf$-values from \cite{LD89}. \cite{scott_sc} also performed calculations with the classical model atmosphere from the MARCS database, which is also used in our work, and obtained a slightly lower value of $\eps{}$(1D) = 3.14 than that for the 3D model, but slightly higher value (by 0.02~dex) than our non-LTE abundance.

In \cite{sc2_gf_pr}, the abundances published by \cite{scott_sc} for individual lines of \ion{Sc}{2} were recalculated with a replacement of $gf$-values to their own values. The obtained average $\eps{}$(PR17) = 3.04 $\pm 0.13$ is consistent with the meteoritic abundance, but differs from all other determinations by a large mean-square error, which is entirely due to uncertainties in $gf$-values obtained by \cite{sc2_gf_pr}.

Thus, with modern atomic data on the oscillator strengths and the hyperfine structure of the \ion{Sc}{2} lines solar abundance exceeds the meteoritic one, $\eps{met}$ = 3.04$\pm$0.03 \citep{lodders21}, by more than 1$\sigma$, regardless of the used spectra of the Sun (as a star or from the center of the disk) and atmospheric models (semi-empirical or theoretical 1D and 3D).

\section{Evolution of scandium content in the Galaxy}\label{sect:trend}

\subsection{Sample of stars, observational material and atmospheric parameters}

\begin{table}
	\centering
	\renewcommand{\arraystretch}{1.0}
	\renewcommand{\tabcolsep}{10pt}
	\caption{LTE and non-LTE (NLTE) abundances of scandium in the sample stars.} 
	\vspace{3mm}	
	\label{tab:stars} 
    \begin{tabular}{rcrccr}
\hline \noalign{\smallskip} 
 HD/BD & \Teff (K)/\lgg /[Fe/H]/$\xi_{t}$(\kms) & $N_l$ & \multicolumn{2}{c}{$\eps{}$} & [Sc/Fe] \\
\cline{4-5}
       &                                        &       & LTE & NLTE & NLTE \\
    	\noalign{\smallskip} \hline \noalign{\smallskip}
    	 \multicolumn{6}{l}{Stellar sample from \cite{lickI}} \\
19373 & 6045 / 4.24 /~~0.10 / 1.2 & 15 &  3.33(0.11) &  3.30(0.12) &   0.08 \\
22484 & 6000 / 4.07 /~~0.01 / 1.1 & 13 &  3.23(0.12) &  3.19(0.09) &   0.06 \\
22879 & 5800 / 4.29 /--0.84 / 1.0 & 10 &  2.49(0.03) &  2.50(0.03) &   0.22 \\
24289 & 5980 / 3.71 /--1.94 / 1.1 & 11 &  1.18(0.06) &  1.28(0.05) &   0.10 \\
30562 & 5900 / 4.08 /~~0.17 / 1.3 & 14 &  3.34(0.09) &  3.30(0.08) &   0.01 \\
30743 & 6450 / 4.20 /--0.44 / 1.8 & 12 &  2.83(0.11) &  2.80(0.11) &   0.12 \\
34411 & 5850 / 4.23 /~~0.01 / 1.2 & 15 &  3.23(0.09) &  3.21(0.09) &   0.08 \\
43318 & 6250 / 3.92 /--0.19 / 1.7 & 13 &  3.06(0.10) &  3.03(0.11) &   0.10 \\
45067 & 5960 / 3.94 /--0.16 / 1.5 & 12 &  3.06(0.11) &  3.02(0.11) &   0.06 \\
45205 & 5790 / 4.08 /--0.87 / 1.1 & 16 &  2.55(0.06) &  2.54(0.05) &   0.29 \\
49933 & 6600 / 4.15 /--0.47 / 1.7 & 12 &  2.79(0.12) &  2.77(0.10) &   0.12 \\
52711 & 5900 / 4.33 /--0.21 / 1.2 & 11 &  3.04(0.07) &  3.02(0.06) &   0.11 \\
58855 & 6410 / 4.32 /--0.29 / 1.6 & 11 &  3.04(0.13) &  3.01(0.11) &   0.18 \\
59374 & 5850 / 4.38 /--0.88 / 1.2 & 10 &  2.49(0.03) &  2.52(0.03) &   0.28 \\
59984 & 5930 / 4.02 /--0.69 / 1.4 & 11 &  2.72(0.11) &  2.72(0.09) &   0.29 \\
62301 & 5840 / 4.09 /--0.70 / 1.3 & 17 &  2.59(0.05) &  2.57(0.05) &   0.15 \\
64090 & 5400 / 4.70 /--1.73 / 0.7 &  9 &  1.46(0.07) &  1.51(0.05) &   0.12 \\
69897 & 6240 / 4.24 /--0.25 / 1.4 & 10 &  2.96(0.09) &  2.95(0.09) &   0.08 \\
74000 & 6225 / 4.13 /--1.97 / 1.3 &  8 &  1.21(0.05) &  1.32(0.04) &   0.17 \\
76932 & 5870 / 4.10 /--0.98 / 1.3 & 15 &  2.39(0.04) &  2.40(0.05) &   0.26 \\
82943 & 5970 / 4.37 /~~0.19 / 1.2 & 16 &  3.33(0.09) &  3.30(0.10) & --0.01 \\
84937 & 6350 / 4.09 /--2.16 / 1.7 & 22 &  1.10(0.06) &  1.20(0.06) &   0.24 \\
89744 & 6280 / 3.97 /~~0.13 / 1.7 & 12 &  3.38(0.08) &  3.35(0.09) &   0.10 \\
90839 & 6195 / 4.38 /--0.18 / 1.4 & 13 &  3.02(0.09) &  3.00(0.09) &   0.06 \\
92855 & 6020 / 4.36 /--0.12 / 1.3 & 10 &  3.11(0.08) &  3.07(0.06) &   0.07 \\
94028 & 5970 / 4.33 /--1.47 / 1.3 & 12 &  1.78(0.07) &  1.85(0.06) &   0.20 \\
99984 & 6190 / 3.72 /--0.38 / 1.8 & 10 &  2.89(0.07) &  2.87(0.07) &   0.13 \\
100563 & 6460 / 4.32 /~~0.06 / 1.6 &  8 &  3.30(0.13) &  3.25(0.10) &   0.07 \\
102870 & 6170 / 4.14 /~~0.11 / 1.5 & 13 &  3.33(0.08) &  3.30(0.09) &   0.07 \\
103095 & 5130 / 4.66 /--1.26 / 0.9 &  9 &  1.92(0.06) &  1.95(0.06) &   0.09 \\
105755 & 5800 / 4.05 /--0.73 / 1.2 & 13 &  2.72(0.08) &  2.72(0.07) &   0.33 \\
106516 & 6300 / 4.44 /--0.73 / 1.5 &  8 &  2.62(0.05) &  2.64(0.06) &   0.25 \\
108177 & 6100 / 4.22 /--1.67 / 1.1 &  6 &  1.53(0.16) &  1.63(0.16) &   0.18 \\
110897 & 5920 / 4.41 /--0.57 / 1.2 & 13 &  2.78(0.08) &  2.78(0.09) &   0.23 \\
114710 & 6090 / 4.47 /~~0.06 / 1.1 & 11 &  3.24(0.08) &  3.22(0.06) &   0.04 \\
115617 & 5490 / 4.40 /--0.10 / 1.1 & 13 &  3.14(0.10) &  3.13(0.10) &   0.11 \\
134088 & 5730 / 4.46 /--0.80 / 1.1 & 11 &  2.61(0.05) &  2.62(0.05) &   0.30 \\
134169 & 5890 / 4.02 /--0.78 / 1.2 & 17 &  2.56(0.06) &  2.55(0.04) &   0.21 \\
    	\noalign{\smallskip}\hline 
\end{tabular}
\end{table}

\begin{table}
	\centering
	\renewcommand{\arraystretch}{1.0}
	\renewcommand{\tabcolsep}{10pt}
	\caption{Table \ref{tab:stars} is continued.} 
	\vspace{3mm}	
	\begin{tabular}{rcrccr}
		\hline \noalign{\smallskip} 
		HD/BD & \Teff (K)/\lgg /[Fe/H]/$\xi_{t}$(\kms) & $N_l$ & \multicolumn{2}{c}{$\eps{}$} & [Sc/Fe] \\
		\cline{4-5}
		&                                        &       & LTE & NLTE & NLTE \\
		\noalign{\smallskip} \hline \noalign{\smallskip} 
138776 & 5650 / 4.30 /~~0.24 / 1.3 & 14 &  3.46(0.10) &  3.45(0.10) &   0.09 \\
140283 & 5780 / 3.70 /--2.46 / 1.6 &  9 &  0.71(0.06) &  0.84(0.06) &   0.18 \\
142091 & 4810 / 3.12 /--0.07 / 1.2 & 12 &  3.12(0.13) &  3.14(0.14) &   0.09 \\
142373 & 5830 / 3.96 /--0.54 / 1.4 & 13 &  2.89(0.10) &  2.88(0.10) &   0.30 \\
+7$^\circ$ 4841 & 6130 / 4.15 /--1.46 / 1.3 & 16 &  1.78(0.05) &  1.85(0.04) &   0.19 \\
+9$^\circ$ 0352 & 6150 / 4.25 /--2.09 / 1.3 &  8 &  1.15(0.06) &  1.30(0.06) &   0.27 \\
+24$^\circ$ 1676 & 6210 / 3.90 /--2.44 / 1.5 &  7 &  0.92(0.07) &  1.06(0.07) &   0.38 \\
+29$^\circ$ 2091 & 5860 / 4.67 /--1.91 / 0.8 &  9 &  1.35(0.10) &  1.46(0.06) &   0.25 \\
+37$^\circ$ 1458 & 5500 / 3.70 /--1.95 / 1.0 & 10 &  1.26(0.11) &  1.34(0.06) &   0.17 \\
+66$^\circ$ 0268 & 5300 / 4.72 /--2.06 / 0.6 &  8 &  1.00(0.09) &  1.09(0.05) &   0.03 \\
--4$^\circ$ 3208 & 6390 / 4.08 /--2.20 / 1.4 &  9 &  1.08(0.04) &  1.19(0.06) &   0.27 \\
--13$^\circ$ 3442 & 6400 / 3.95 /--2.62 / 1.4 &  7 &  0.81(0.07) &  0.93(0.07) &   0.43 \\
G090-003 & 6007 / 3.90 /--2.04 / 1.3 & 10 &  1.22(0.10) &  1.32(0.09) &   0.24 \\
 \multicolumn{6}{l}{Stars from \cite{mash2003}} \\
 31128 &  5980 / 4.42 / --1.53 / 1.2 &  8 &  1.74(0.04) &   1.81 (0.05)  &   0.22  \\
 97320 &  6110 / 4.26 / --1.18 / 1.4 &  8 &  2.20(0.05) &   2.23 (0.06)  &   0.29  \\
102200 &  6115 / 4.29 / --1.19 / 1.4 &  8 &  2.01(0.03) &   2.04 (0.05)  &   0.11  \\
193901 &  5780 / 4.46 / --1.08 / 0.9 &  6 &  2.14(0.06) &   2.18 (0.06)  &   0.14   \\    	 
298986 &  6130 / 4.26 / --1.36 / 1.4 &  8 &  1.87(0.06) &   1.92 (0.05)  &   0.16   \\
    	\noalign{\smallskip}\hline 
\multicolumn{6}{l}{{\bf Note.} Numbers in parentheses are the abundance errors $\sigma$.}  \\  	
	\end{tabular}
\end{table}

We use the stellar sample from our previous works, \cite{lickI} and \cite{Zhao2016}, in total 51 stars in the metallicity range $-2.62 \le$ [Fe/H] $\le 0.24$. \cite{lickI} provide all information on spectral observations and their reduction. Here, we briefly point out that for 47 stars the spectra were obtained in Lick Observatory (USA), on a 3-m telescope with a Hamilton echelle spectrograph; R $\simeq$ 60\,000, the signal-to-noise ratio (S/N) at a wavelength of 5500~\AA exceeds 100. The spectra of HD~84937 and HD~140283 with R $\simeq$ 80\,000 are taken from the UVESPOP archive (Bagnulo et al., 2003). Spectra of two more stars were obtained with the Canadian-French-Hawaiian Telescope (CFHT) with the ESPaDOnS spectrograph; R $\simeq$ 60\,000.

The atmospheric parameters are taken from \cite{lickI}. The effective temperatures and surface gravities were determined using several methods, in particular, the infrared flux method, the distance based method using Hipparcos trigonometric parallaxes, and the spectroscopic method based on the non-LTE analysis of lines of \ion{Fe}{1} and \ion{Fe}{2}. The latter method also gave the iron abundance [Fe/H] and microturbulent velocity. The stellar sample is homogeneous in temperature and luminosity, which are close to solar values: 5400~K $\le$ \Teff\ $\le$ 6600~K and 3.70 $\le$ \lgg\ $\le$ 4.72. Exception is a cool giant HD~142091 with \Teff\ = 4810~K and \lgg\ = 3.12, which is still far from bringing nucleosynthesis products to the surface. The stars and their atmospheric parameters are listed in Table~\ref{tab:stars}.

To improve statistics in the range $-1.5 <$ [Fe/H] $< -1$, the sample was complemented with five stars from \cite{mash2003}. For four stars of them, the spectra were obtained at the European Southern Observatory, using the VLT2 telescope with the UVES echelle spectrograph; R $\simeq$ 80\,000, everywhere S/N $> 100$. Spectrum of HD~193901 was obtained by K.~Fuhrmann at the Spanish-German observatory Calar-Alto (Spain) with a 2.2-m telescope with a FOCES echelle spectrograph; R $\simeq$ 60\,000, S/N $> 100$ at wavelengths greater 4500~\AA.

For these five stars, the effective temperatures were determined by \cite{mash2003} from H$_{\alpha}$ and H$_{\beta}$ line wings fitting. The surface gravities were revised in this paper using the Gaia EDR3 parallaxes \citep{gaia_edr3} and the apparent visual magnitudes, masses, and \Teff\ from \cite{mash2003}. All the stars are close by, so their \lgg\ changed by no more than 0.09~dex, compared with the results of \cite{mash2003}, which were based on the Hipparcos parallaxes. We have revised [Fe/H] and $\xi_t$ from analysis of the \ion{Fe}{2} lines. In fact, the microturbulent velocities have not changed. These five stars are also of solar type with \Teff\ ranging from 5780~K to 6130~K and \lgg\ from 4.26 to 4.46. The results are shown in Table~\ref{tab:stars}.

\subsection{Scandium Abundances and Galactic Trend [Sc/Fe]}

The scandium abundances were determined by the synthetic spectrum method, as for the reference stars in Section~\ref{sect:sun}. The results are presented in Table~\ref{tab:stars}. The [Sc/Fe] values are calculated with the solar non-LTE abundance determined in this paper: $\eps{NLTE}$ = 3.12.

For most of the stars, the rms error of the non-LTE abundance is slightly less or the same as in the case of LTE, and does not exceed 0.11~dex. The largest abundance scatter for individual lines was obtained for HD~108177 ($\sigma$ = 0.16~dex) and HD~142091 ($\sigma$ = 0.14~dex). In the first star with [Fe/H] = $-1.67$ we were able to measure only 6 lines due to the lack of blue ($\lambda <$ 4500~\AA) part of the spectrum and weaknesses of all lines with $\lambda \ge$ 5669~\AA. The second star is the coolest (\Teff\ = 4810~K) in the sample, and the resulting low abundance from the 4314 and 4400~\AA\ lines may be due to an overestimation of the contribution of molecular lines.

\begin{figure}  
	\centering
	\includegraphics[width=0.45\columnwidth,clip]{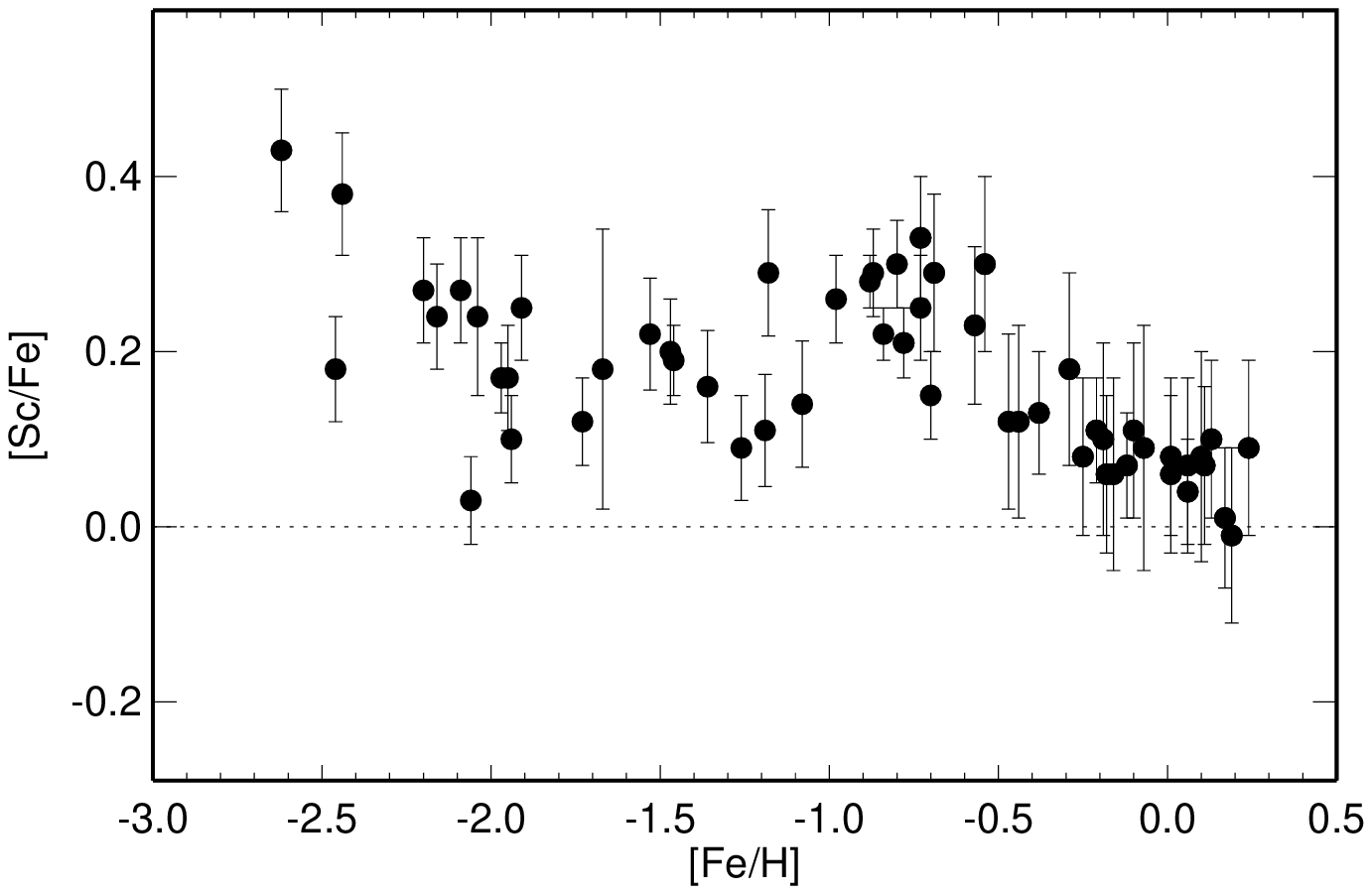}
	\includegraphics[width=0.45\columnwidth,clip]{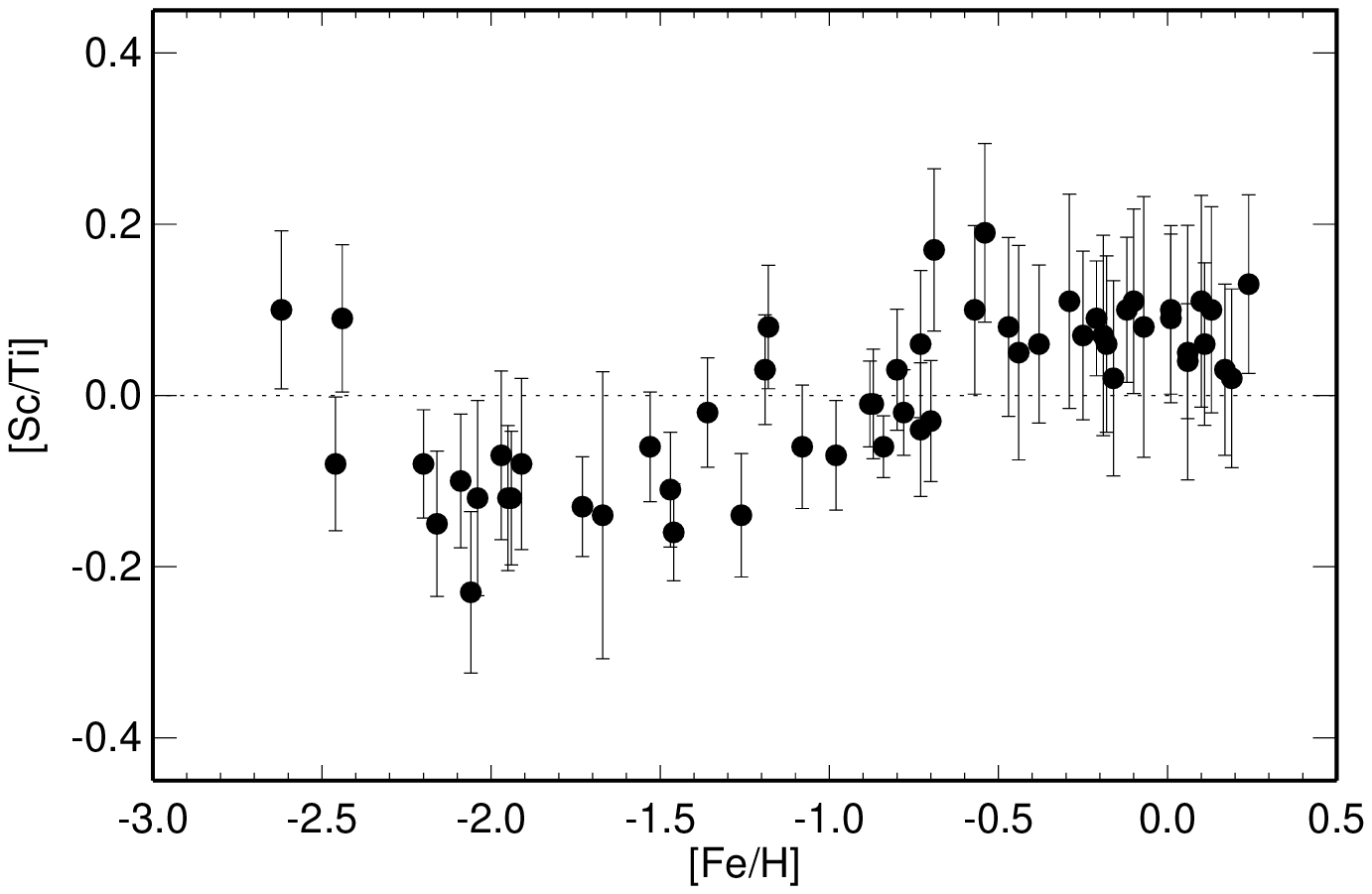}
	\caption{[Sc/Fe] and [Sc/Ti] as a function of [Fe/H] for 56 stars in the solar neighborhood. All results are based on the non-LTE calculations.}
	\label{fig:sc_trend}
\end{figure}

In our sample, seven stars have metallicity, close to the solar one (within 0.10~dex), and they all reveal a slight enhancement of Sc relative to the solar value, with the average [Sc/H] = 0.08. Similar results were obtained by \cite{Zhao2016}, who applied a line-by-line differential approach (for each individual line, its abundance is compared with abundance of its solar counterpart) and the non-LTE method of \cite{2008A&A...481..489Z}. For the same seven stars they obtained the average [Sc/H] = 0. 0.4. Thus, not only the Sun but also the solar-type stars in the solar neighborhood have the higher Sc abundance compared with the meteoritic one.

The evolution of the [Sc/Fe] ratio with the change in [Fe/H] is shown in Fig.~\ref{fig:sc_trend}. In the [Fe/H] $> -1$ range, [Sc/Fe] decreases with increasing [Fe/H], and this is similar to the behavior of the $\alpha$-process elements O, Mg, Si, Ca, which are synthesized in SNeII and for which [$\alpha$/Fe] falls in the same range [Fe/H], due to the onset of Fe production in Type Ia supernovae (SNeIa). For lower iron abundances, $\alpha$-elements reveal a plateau at the level of [$\alpha$/Fe] $\simeq$ 0.3-0.4 (see Fig.~8 and references in \cite{Zhao2016}). Scandium in this range is also observed in excess relative to iron, but the excess is less, at the level of $\sim$ 0.2~dex, and there are significant deviations from this value. The two stars, BD~+24$^\circ$~1676 and BD~$-13^\circ$~3442, have [Sc/Fe] $\sim$ 0.4, while [Sc/Fe] $\sim$ 0 for BD~+66$^\circ$~0268.

The second panel of Fig.~\ref{fig:sc_trend} shows the [Sc/Ti] ratios. There is a hint that they grow from [Sc/Ti] $\sim -0.1$ to [Sc/Ti] $\sim +0.1$ with increasing [Fe/H]. The exceptions are again BD~+24$^\circ$~1676 and BD~$-13^\circ$~3442. The Ti non-LTE abundances are taken from \cite{Zhao2016} and  \cite{sitnova_zn}.

The origin of Sc is poorly understood, and existing models of nucleosynthesis suffer from too low production of this element (Kobayashi et al. 2020). We do not plot the predictions of the Galactic chemical evolution model on Fig.~\ref{fig:sc_trend}, since the curve would pass well below the observed points. For example, the model of Kobayashi et al. (2020) predicts [Sc/Fe] = $-1.15, -0.8$, and $-0.9$ for [Fe/H] =  $-2.5, -1.0$, and 0.0, respectively. For titanium, the theoretical models also cannot reproduce observations where Ti behaves like an $\alpha$-element (for example, Fig.~8 in \cite{Zhao2016}).

Apparently, the problem of scandium synthesis should be solved together with the problem of titanium synthesis. We hope that the Sc abundances obtained in this study for a wide metallicity range will push a new impetus to ideas about the origin of scandium.

{\it Comparison with the results of Zhao et al. (2016, hereinafter ZMY16).} The obtained [Sc/Fe] -- [Fe/H] trend is consistent with ZMY16 data for [Fe/H] $> -0.9$, but we obtained a higher Sc abundance for the lower metallicity stars. For example, for HD~84937 [Sc/Fe](non-LTE) = 0.24 in this work, but 0.10 in ZMY16, and this is not related to different estimates of the non-LTE effects. The difference between non-LTE and LTE is 0.10~dex in this study and 0.12~dex in ZMY16. The most dramatic difference between this work and ZMY16 is obtained for BD~$-13^\circ$~3442: [Sc/Fe](non-LTE) = 0.43 and 0.06, respectively. The difference between non-LTE and LTE is 0.12~dex in this work, and it is even larger, of 0.16~dex, in ZMY16. We believe that the main reason for the discrepancies lies with the list of lines used to determine the Sc abundances. We analyzed the \ion{Sc}{2} 4246, 4314, 4320, 4400\,\AA\ lines which remain strong even in the VMP stars, while these lines were excluded in a line-by-line differential approach adopted by ZMY16, since they cannot be analyzed in the solar spectrum. The \ion{Sc}{2} lines in the list of ZMY16 are very weak in stars with [Fe/H] $< -2$. As a result, for BD~$-13^\circ$~3442 our Sc average abundance is based on seven lines, while ZMY16 used three weak lines.

Our results agree with the data of \cite{Reggiani_sc} in the overlapping metallicity range $-2.6 <$ [Fe/H] $< -1.5$. For 23 stars Reggiani et al. (2017) obtained an average of [Sc/Fe] = 0.31.

\section{Conclusions}\label{conclusions}

We constructed a new model atom of \ion{Sc}{2} using the most up-to-date atomic data. Due to the lack of accurate calculations of the \ion{Sc}{2} + \ion{H}{1} collisions, they are treated using the approximate formula of \cite{1984A&A...130..319S}. Our non-LTE calculations with and without hydrogenic collisions show that their influence on the final results is small even for VMP stars. For example, for HD~84937 ([Fe/H] = $-2.16$) the abundance difference between these two options is $-0.02$~dex (positive non-LTE corrections are larger when only collisions with electrons are taken into account).

The developed method was tested by analyzing the \ion{Sc}{2} lines in spectra of the reference stars the Sun, Procyon, and HD~84937. For each star, taking into account the departures from LTE reduces the scatter in the abundance determined from different lines, including lines in the UV and visible ranges for HD~84937, and reduces the abundance error.

Solar scandium non-LTE abundance $\eps{NLTE}$ = 3.12$\pm$0.05 is 0.02~dex less than the LTE value and 0.05~dex and 0.04~dex less than the values obtained by Scott et al. (2015) and Lawler et al. (2019), respectively. And yet it exceeds the meteoritic value, $\eps{met}$ = 3.04$\pm$0.03 \citep{lodders21}, by more than 1$\sigma$. In Procyon, the scandium non-LTE abundance is consistent with the meteoritic one.

The developed method was applied to determine the scandium abundances of 56 solar-type stars in the $-2.62 \le$ [Fe/H] $\le 0.24$ metallicity range using high-resolution spectra.

Seven stars have a close-to-solar metallicity ($-0.10 \le$ [Fe/H] $\le 0.10$), and they all have the scandium abundance, which is higher than the meteoritic one and also the solar one, by 0.08~dex, on average. This makes the question of what is the cosmic abundance of scandium even more relevant and requires further research by increasing the star statistics and analysing correlations of the Sc abundances with various stellar parameters, such as star's mass, age, etc.

In the [Fe/H] $< -1$ range, scandium is enhanced relative to iron with [Sc/Fe] $\sim$ 0.2~dex, although the three stars reveal a significant deviation from this value. For higher [Fe/H], the [Sc/Fe] ratio drops to a value close to solar. This behavior indicates the synthesis of scandium in massive stars and resembles the Galactic trend [$\alpha$/Fe] -- [Fe/H], although in the [Fe/H] $< -1$ region the $\alpha$-elements have higher enhancement relative to iron.

Our results demonstrate a correlation between abundances of scandium and titanium. The origin of both is still unclear. We hope that the scandium abundances obtained for a wide range of metallicities will push a new impetus to ideas about Sc (and Ti ?) origin.

L.M. acknowledges the support of Ministry of Science and
Higher Education of the Russian Federation under the grant 075-15-2020-780 (N13.1902.21.0039).

\end{document}